\documentclass{article}

\usepackage{arxiv}

\usepackage[utf8]{inputenc} 
\usepackage[T1]{fontenc}    
\usepackage{hyperref}       
\usepackage{url}            
\usepackage{booktabs}       
\usepackage{amsfonts}       
\usepackage{nicefrac}       
\usepackage{microtype}      
\usepackage{graphicx}
\usepackage[numbers,sort&compress]{natbib}
\usepackage{doi}
\usepackage{makecell}

\title{AutoIQ: An Ensemble Framework for Automatic Assessment of Geometric Distortion in Prostate Diffusion-Weighted Imaging}

\date{} 					

\author{
Haoran Sun$^{1,2}$,
Lixia Wang$^{1}$,
Yin-Chen Hsu$^{1}$,
Hsu-Lei Lee$^{1}$,
Chang Gao$^{3}$,
Fei Han$^{3}$,\\
\textbf{
Robert Grimm$^{4}$,
Vibhas Deshpande$^{5}$,
Ziyang Long$^{1,2}$,
Hsin-Jung Yang$^{1}$,
Rola Saouaf$^{6}$,
Alessandro D'Agnolo$^{7}$,}\\
\textbf{
Timothy Daskivich$^{8}$,
Hyung Kim$^{8}$,
Debiao Li$^{1,2}$,
Yibin Xie$^{1}$\thanks{Corresponding author. Email: \texttt{Yibin.Xie@cshs.org}.}}
\\
\\
$^{1}$Biomedical Imaging Research Institute, Cedars-Sinai Medical Center, Los Angeles, CA, USA\\
$^{2}$Department of Bioengineering, University of California, Los Angeles, CA, USA\\
$^{3}$Siemens Medical Solutions USA Inc., Los Angeles, CA, USA\\
$^{4}$Siemens Healthineers AG, Forchheim, Germany\\
$^{5}$Siemens Medical Solutions USA Inc., Austin, TX, USA\\
$^{6}$Department of Imaging, Cedars-Sinai Medical Center, Los Angeles, CA, USA\\
$^{7}$Department of Nuclear Medicine, Cedars-Sinai Medical Center, Los Angeles, CA, USA\\
$^{8}$Department of Urology, Cedars-Sinai Medical Center, Los Angeles, CA, USA
}

\renewcommand{\headeright}{ }
\renewcommand{\undertitle}{ }
\renewcommand{\shorttitle}{\textit{AutoIQ}}

\begin{document}
\maketitle

\begin{abstract}
Geometric distortion in prostate diffusion-weighted imaging (DWI) can impair lesion localization and reduce the reliability of MRI-based clinical assessment. We propose AutoIQ, an ensemble machine learning framework for automatic quantification and classification of DWI geometric distortion severity. A total of 140 retrospective prostate biparametric MRI examinations were analyzed, including 33 scans with severe distortion requiring repeat acquisition and 107 scans with acceptable distortion based on expert radiologist assessment. AutoIQ combines two complementary distortion quantification strategies: a segmentation-based method measuring prostate boundary mismatch between T$_2$-weighted imaging (T2WI) and DWI, and a registration-based method estimating deformation magnitude after DWI-to-T2WI alignment. The resulting distortion scores were used to train individual classifiers and a logistic-regression ensemble model. Both computational methods significantly differentiated severe from acceptable distortion cases (\(p < 0.001\)). On an independent test set, the ensemble model achieved an accuracy of 0.95, F1-score of 0.93, and AUC of 0.98, outperforming individual models. These results suggest that AutoIQ can provide automated, quantitative quality assessment for prostate DWI and may help identify scans that require repeat acquisition.
\end{abstract}


\section{Introduction}
Prostate cancer (PCa) is the most common cancer among men, with an estimated 313,780 new cases and 35,770 deaths projected in the United States for 2025 \citep{Siegel2025Cancer}. Multiparametric magnetic resonance imaging (mpMRI) has become an integral tool for prostate cancer management, aiding in targeted biopsy, staging, and treatment planning \citep{Scheenen2015Multiparametric,Delongchamps2011Multiparametric}.

Within mpMRI, diffusion-weighted imaging (DWI) plays a crucial role, demonstrating high diagnostic performance in detecting prostate cancer and assessing tumor aggressiveness \citep{Turkbey2011Is,desouza2007Magnetic,Vargas2011Diffusion,Yagci2011value}. The significance of DWI is emphasized in the Prostate Imaging Reporting and Data System (PI-RADS) guidelines, which identify it as the primary imaging component for the evaluation of clinically significant prostate cancer, particularly in the peripheral zone \citep{Turkbey2019Prostate}. In addition to the anatomical information provided by high-resolution T$_2$-weighted imaging (T2WI), the increasing use of biparametric MRI in clinical practice, which excludes dynamic contrast-enhanced imaging and relies on DWI as the sole functional sequence, further underscores the importance of DWI \citep{Mahajan2022Evaluation,Brembilla2022Diagnostic}.

However, the single-shot echo-planar imaging (SS-EPI) technique, commonly used in clinical examinations, is highly susceptible to geometric distortion due to its sensitivity to magnetic field inhomogeneities \citep{Le2006Artifacts,Poustchi2001Principles}. Severe distortions not only compromise quantitative apparent diffusion coefficient (ADC) measurements but also result in spatial mismatches that affect MRI-guided biopsy and radiation therapy planning, potentially making the MRI examination uninterpretable~\citep{Nketiah2018Geometric,Weygand2016Spatial,Stocker2018Image}.

Quantitative assessment of geometric distortion severity is rarely available in routine clinical practice. Technologists and radiologists currently rely on subjective visual inspection to judge whether distortion is acceptable, a process that is time-consuming and observer-dependent. Automated, objective quantification of image distortion could provide real-time feedback to MRI technologists at the point of scanning, potentially reducing the need for patient recalls and improving scan quality.

Several prior studies have proposed computational methods for estimating geometric distortion between DWI and distortion-free anatomical reference images such as T2WI \citep{Stocker2018Image,Gill2017method,Donato2014Geometric}. While these methods provide quantitative distortion metrics, they remain stand-alone quality assessment tools and do not predict whether the observed distortion would impair radiologic interpretation.

In this study, we propose AutoIQ (Automated Assessment of Image Quality), an ensemble machine learning framework designed to automatically assess distortion severity in prostate DWI. The method integrates two complementary computational algorithms to quantify geometric distortion, producing distortion factors that serve as inputs to an automated ensemble classifier trained using radiologist assessments as the reference standard. AutoIQ can assist MRI technologists in determining whether a DWI scan is acceptable for interpretation or likely to require repeat acquisition, supporting the delivery of clinically acceptable prostate MRI and helping maintain consistent diagnostic quality.

\section{Materials and methods}
\subsection{Study dataset and data preparation}
This retrospective study included 140 prostate biparametric MRI examinations acquired between January 2018 and December 2022 using a 3T clinical scanner (Biograph mMR; Siemens Healthineers, Erlangen, Germany). All patients were male, with a median age of 69 years (interquartile range, 63.5--73 years). Only examinations with all required image sequences available were included. This study was approved by the Institutional Review Board, and the requirement for informed consent was waived. A subset of the MRI dataset used in this study was previously included in a separate publication focused on retrospective T$_2$ mapping of the prostate \citep{Sun2025Retrospectively}. The current study represents a distinct investigation focused on automated quantification of diffusion-weighted imaging distortion, with no overlap in analytical objectives, methods, or reported results.

Each case included T2WI and DWI sequences. Detailed imaging parameters are provided in Table~\ref{tab:protocol}. Whole-gland prostate segmentations on both modalities were automatically generated using research software (MR Prostate AI; Siemens Healthineers AG, Forchheim, Germany).

\begin{table}[htbp]
\centering
\caption{Imaging protocol parameters for T2WI and DWI.}
\label{tab:protocol}
\footnotesize
\setlength{\tabcolsep}{5pt}
\renewcommand{\arraystretch}{1.25}
\begin{tabular}{lcc}
\toprule
Parameter & \makecell{T2WI\\(TSE / TIRM)} & \makecell{DWI\\(SS-EPI)} \\
\midrule
TE (ms) & 132 / 80 & 95 / 73 \\
TR (ms) & 4000 / 5520 & 5000 / 4500 \\
Flip angle (\(^{\circ}\)) & 158 / 120 & 90 / 90 \\
Number of slices & 30 / 23 & 30 / 23 \\
Slice thickness (mm) & 3 / 3 & 3 / 3 \\
Resolution (mm\(^2\)) 
& \makecell{\(0.7 \times 0.7\) /\\ \(1.17 \times 1.17\)} 
& \makecell{\(2.0 \times 2.0\) /\\ \(2.3438 \times 2.3438\)} \\
FOV (mm\(^2\)) 
& \makecell{\(170 \times 170\) /\\ \(300 \times 300\)} 
& \makecell{\(200 \times 200\) /\\ \(300 \times 300\)} \\
\(b\)-value (s/mm\(^2\)) & --- & 50, 800, 1400 \\
Scan time (min) & 4.5 / 1.1 & 5 / 4.43 \\
\bottomrule
\end{tabular}

\vspace{0.3em}
\begin{minipage}{0.92\linewidth}
\footnotesize
T2WI = T2-weighted imaging; DWI = diffusion-weighted imaging; TSE = turbo spin echo; TIRM = turbo inversion recovery magnitude; SS-EPI = single-shot echo-planar imaging; FOV = field of view.
\end{minipage}
\end{table}

A board-certified radiologist with 25 years of experience independently reviewed each case and categorized distortion severity into two groups: severe distortion requiring repeat acquisition (\(n = 33\)) and acceptable distortion sufficient for clinical interpretation (\(n = 107\)). Distortion grading was performed primarily on axial low b-value DWI (\(b = 50~\mathrm{s/mm^2}\)) with reference to T2WI and ADC maps. This labeling served as the reference standard for model development.

Distortion severity was classified based on the clinical interpretability of DWI. Acceptable distortion, or Rank 0, referred to scans in which distortion may be present but did not substantially impair anatomical interpretation or lesion localization. Characteristics included minor boundary mismatch between T2WI and DWI, mild rectal wall displacement without affecting prostate contour assessment, limited geometric warping not impacting lesion localization, and preserved overall gland shape. Severe distortion, or Rank 1, referred to scans in which distortion was considered clinically relevant and likely to compromise interpretation. Characteristics included visible geometric warping of the peripheral zone boundary, apparent displacement of the prostate capsule relative to T2WI, lesion misregistration affecting localization accuracy, signal stretching, compression, or pile-up resulting in altered gland contour, and significant anatomical mismatch that may impact biopsy targeting or radiation planning. The geometric assessment principles were consistent with distortion-related quality considerations described in the PI-QUAL framework for prostate MRI, although formal PI-QUAL scoring was not performed.

To evaluate labeling reliability, a second board-certified radiologist independently reviewed a randomly selected subset of 35 cases, blinded to the initial assessment. Inter-reader agreement demonstrated excellent concordance (Cohen's \(\kappa = 0.94\), 95\% CI: 0.77--1.00) with 97.1\% overall agreement, supporting the robustness of the reference standard.

\begin{figure}[htbp]
    \centering
    \includegraphics[width=\linewidth]{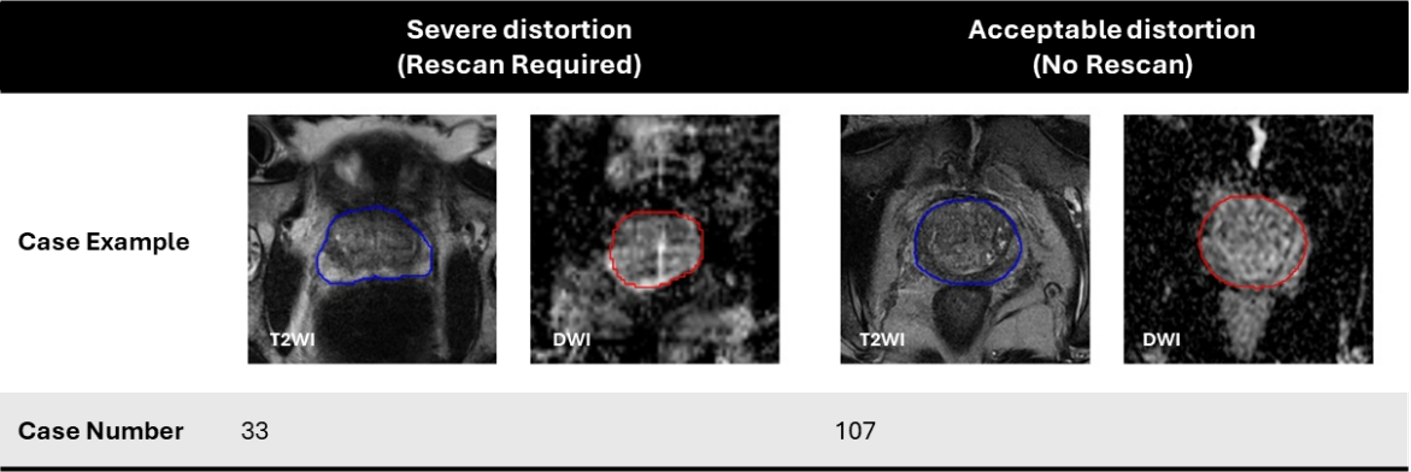}
    \caption{Representative examples of prostate biparametric MRI cases with different levels of geometric distortion severity. Axial T2WI and DWI (\(b = 50~\mathrm{s/mm^2}\)) are shown with prostate gland segmentations overlaid. The left pair illustrates a case with severe distortion requiring repeat acquisition, while the right pair shows a case with acceptable distortion requiring no rescan.}
    \label{fig:fig1}
\end{figure}

The dataset also included eight cases with bilateral metallic implants to ensure representation of more extreme distortion scenarios. Representative examples of severe and acceptable distortion levels are shown in Figure~\ref{fig:fig1}.

\subsection{Computational methods for distortion quantification}
Two complementary computational methods were developed to quantify geometric distortion in prostate DWI: a segmentation-based method that relies on anatomical boundary differences between T2WI and DWI, and a registration-based method that captures voxel-wise deformation through image registration. The DWI image used for distortion analysis was the low b-value image (\(b = 50~\mathrm{s/mm^2}\)). Both methods were implemented in MATLAB 2023a.

\subsubsection{Segmentation-based method}
The segmentation-based method was adapted from a previously published technique by Gill et al., which maps and quantifies whole-organ distortion by comparing gland contours in polar coordinates across T2WI and DWI \citep{Gill2017method}. The detailed computational steps are illustrated in Figure~\ref{fig:fig2}(A). Prostate glands were segmented during data preparation on both image modalities. For each slice, the contours from T2WI and DWI were aligned by overlapping their geometric centers. The radial distance between corresponding boundary points was calculated in polar coordinates, and a distortion factor was computed as the normalized sum of squared radial displacements between the two contours. A distortion factor was computed for each slice and then averaged across all slices to generate the scan-level mean distortion score.

\subsubsection{Registration-based method}
\begin{figure}[htbp]
    \centering
    \includegraphics[width=\linewidth]{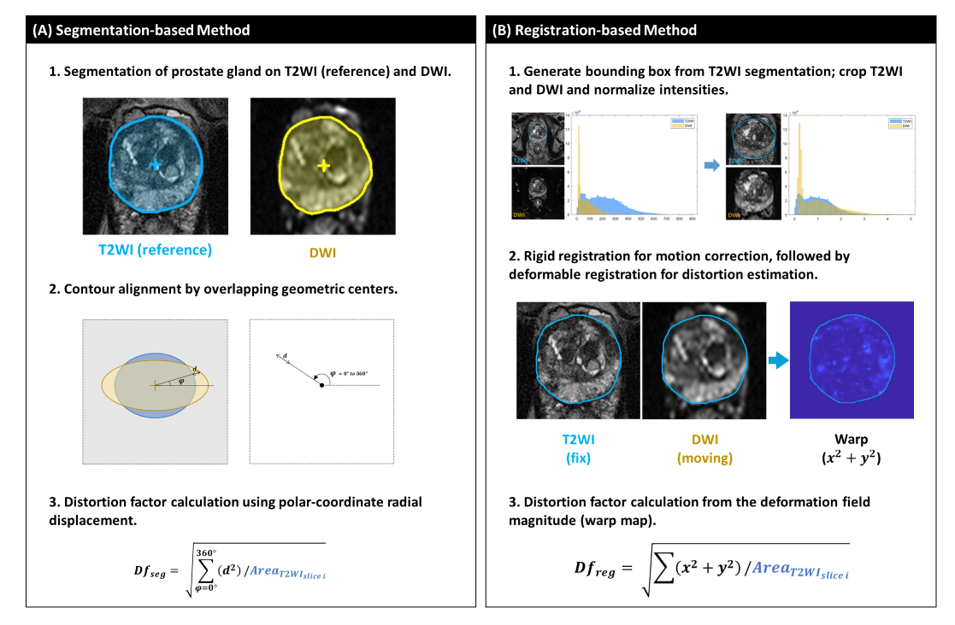}
    \caption{Illustration of the two computational methods used for distortion quantification in prostate DWI. (A) Segmentation-based method: Prostate glands are segmented on both T2WI (reference) and DWI. After aligning the segmented contours by overlapping their geometric centers, slice-wise distortion factors are calculated from radial displacements in polar coordinates, normalized by the T2WI gland area. (B) Registration-based method: A bounding box is generated from the T2WI segmentation and used to crop and normalize both T2WI and DWI. Rigid registration is applied for motion correction, followed by deformable registration to estimate distortion. The distortion factor is computed from the magnitude of the deformation field (warp map), normalized by the T2WI slice area.}
    \label{fig:fig2}
\end{figure}

The registration-based method quantified distortion based on deformation fields generated by non-rigid registration of DWI to T2WI, as illustrated in Figure~\ref{fig:fig2}(B). First, a bounding box was generated from the T2WI prostate segmentation and used to crop both T2WI and DWI. The cropped images were intensity-normalized to minimize modality-related contrast differences. A two-step registration process was performed: rigid registration was applied to correct for global patient movement, followed by deformable registration to estimate geometric distortion. The resulting deformation field, or warp map, was used to compute the distortion factor, defined as the mean magnitude of voxel displacements within the prostate region on each slice. These per-slice distortion factors were then averaged to obtain a scan-level mean distortion score.

\subsection{Ensemble machine learning framework}
\begin{figure}[htbp]
    \centering
    \includegraphics[width=\linewidth]{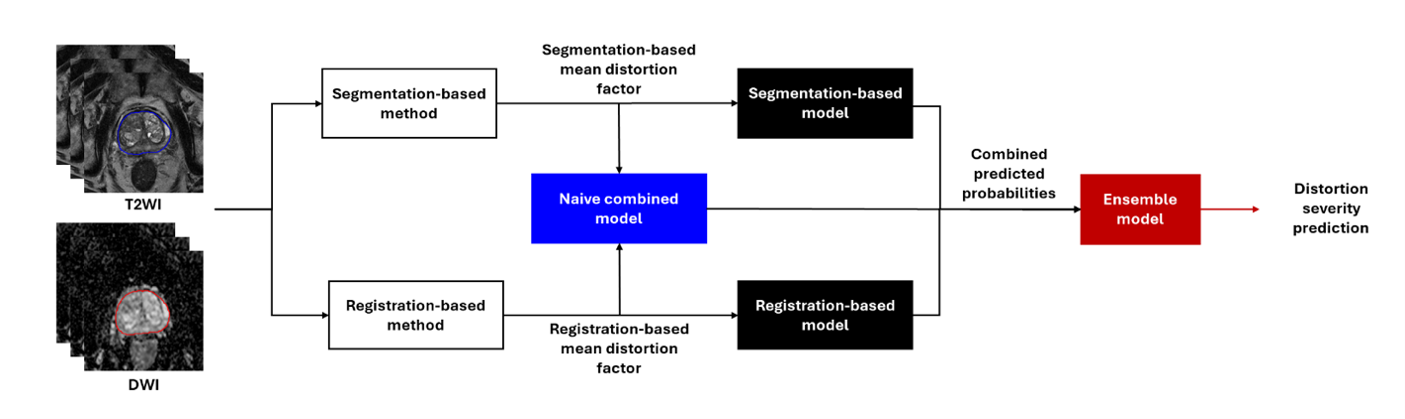}
    \caption{Pipeline of the AutoIQ ensemble machine learning framework for automatic distortion severity assessment in prostate DWI. Segmentation-based and registration-based computational methods were used to calculate distortion factors, respectively, which were independently used to train two base classifiers. A third model (naive-combined) was trained using concatenated distortion factors from both methods. The ensemble model integrates the probability outputs from all three classifiers to generate the final distortion severity group prediction.}
    \label{fig:fig3}
\end{figure}

As shown in Figure~\ref{fig:fig3}, an ensemble machine learning framework was developed to predict the distortion severity of prostate DWI. The model classified each DWI scan into one of two categories based on the radiologist-provided labels: severe distortion, requiring rescan, or acceptable distortion, requiring no rescan. Three base classifiers were trained separately using support vector machines (SVMs): one using distortion factors derived from the segmentation-based computational method, one using distortion factors from the registration-based computational method, and a naive combined model that used concatenated features from both methods. To leverage the complementary strengths of these models, an ensemble classifier was constructed using logistic regression, which integrated the probability outputs from the three SVM base models to generate a final prediction. This soft-voting strategy enabled the ensemble model to learn an optimal combination of anatomical precision from segmentation-based features, which focus on boundary alignment; global robustness from registration-based features, which capture broad spatial deformation patterns; and joint feature representation from the naive model, resulting in improved classification performance. Model development was conducted in Python 3.8.18 using the scikit-learn library, version 1.3.2.

\subsection{Statistical analysis and model evaluation}
To evaluate whether the computed distortion factors could distinguish between radiologist-assigned distortion severity levels, the mean distortion factor from each method, segmentation-based and registration-based, was calculated per case. A non-parametric Mann--Whitney U test was used to compare the distributions of mean distortion factors between the two severity groups: severe distortion, requiring rescan, and acceptable distortion, requiring no rescan. Statistical significance was defined as \(p < 0.05\).

Four classification models were evaluated: segmentation-based model, registration-based model, naive combined model, and the proposed ensemble model. The full dataset of 140 cases was randomly divided into a training set (\(n = 85\)) and an independent testing set (\(n = 55\), approximately 40\%), using stratified sampling to preserve class balance between severe and acceptable distortion categories. Model performance was first assessed on the training set using five-fold cross-validation and subsequently evaluated on the independent testing set to examine generalizability. Classification performance was evaluated using six metrics: accuracy, recall, specificity, precision, F1-score, and the area under the receiver operating characteristic curve (AUC).

Code for the AutoIQ framework is available at \url{https://github.com/sunhPG/AutoIQ-prostateDWI}. The MRI data used in this study are not publicly available due to patient privacy restrictions but may be shared upon reasonable request and institutional approval.

\section{Results}
\subsection{Quantitative assessment of distortion factors generated by computational methods}
Two computational methods, segmentation-based and registration-based, were applied to all 140 cases to compute distortion factors for each scan. Quantitative assessments were then performed to evaluate their ability to distinguish between clinically acceptable and unacceptable levels of distortion, as labeled by expert radiologists.

\begin{figure}[htbp]
    \centering
    \includegraphics[width=\linewidth]{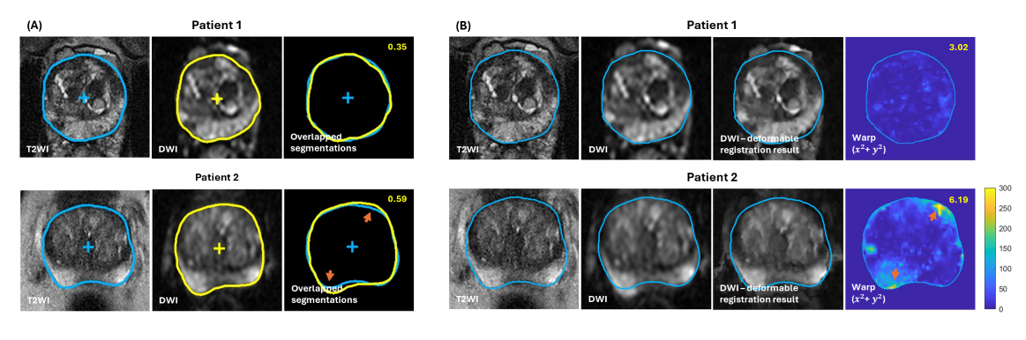}
    \caption{Results of the two distortion quantification methods for two representative patients with differing levels of distortion severity. Patient 1 exhibits lower geometric distortion compared to Patient 2. (A) Segmentation-based method: prostate gland segmentations on T2WI and DWI (\(b = 50\)) are shown in blue and yellow, respectively. The geometric centers of the segmentations are marked with cross symbols. The third column shows the overlaid segmentations, with the distortion factor of the slice displayed in the top right. (B) Registration-based method: The prostate gland segmentation from the T2WI (blue contour) is used to crop all images into a bounding box. Columns show the original T2WI, DWI, deformable registration result, and the resulting warp map. Hotter colors in the warp map (yellow regions) indicate larger displacements as highlighted by orange arrows.}
    \label{fig:fig4}
\end{figure}

Figure~\ref{fig:fig4} illustrates representative results from two patients exhibiting different degrees of geometric distortion. In the segmentation-based method (Figure~\ref{fig:fig4}(A)), prostate gland segmentations are overlaid in blue on T2WI and in yellow on DWI (\(b = 50~\mathrm{s/mm^2}\)). Cross marks denote the geometric centers of the prostate gland in each modality. In the case with severe distortion (Patient 2), arrows highlight a sheared DWI segmentation that does not align with the T2WI outline. The computed distortion factor was lower for Patient 1 (0.35) and higher for Patient 2 (0.59), consistent with their respective distortion levels. In the registration-based method (Figure~\ref{fig:fig4}(B)), the T2WI segmentation, shown as a blue contour, is used to crop the images to the bounding box surrounding the prostate. Warp maps reveal high-intensity regions indicating localized distortion in DWI. Patient 2 exhibited higher deformation intensity, especially in two distinct regions, resulting in a distortion factor of 6.19, compared with 3.02 in Patient 1, again consistent with the distortion severity.

\begin{figure}[htbp]
    \centering
    \includegraphics[width=0.9\linewidth]{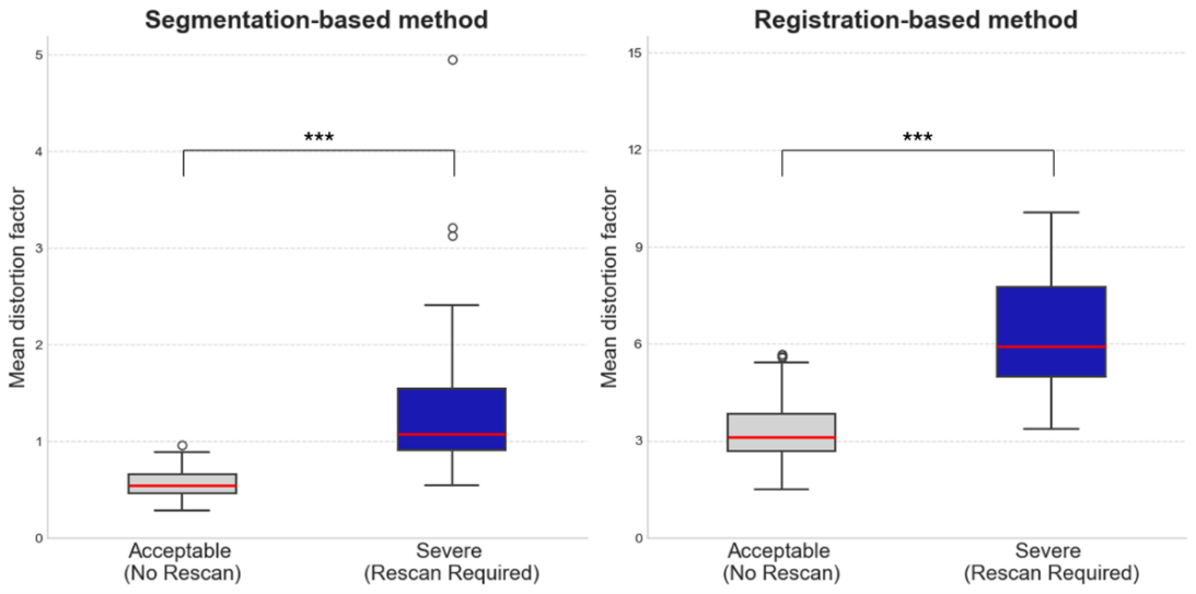}
    \caption{Box plots show the distribution of mean distortion factors for the segmentation-based (left) and registration-based (right) methods, stratified by radiologist-assessed distortion severity. ``Acceptable'' refers to scans that did not require a rescan, while ``Severe'' refers to scans that required repeat acquisition. Boxes represent the interquartile range (IQR), red horizontal lines indicate the median, and whiskers extend to \(1.5 \times \mathrm{IQR}\). Outliers are shown as white circles. A statistically significant difference between the two severity groups was observed for both methods using the Mann--Whitney U test (\(***p < 0.001\)).}
    \label{fig:fig5}
\end{figure}

To quantitatively assess the ability of the methods to distinguish between distortion levels, the mean distortion factor was computed for each case and compared across radiologist-assigned severity categories: acceptable, requiring no rescan, and severe, requiring repeat acquisition. As shown in Figure~\ref{fig:fig5}, mean distortion values were higher in severe distortion cases than in acceptable cases. Both methods produced measurements that strongly aligned with radiologist assessments. Statistically significant differences were observed between the severity classes using the Mann--Whitney U test (\(p < 0.001\)).

These results demonstrate that both computational methods can quantitatively assess distortion severity, providing distortion scores that significantly differentiate between acceptable and clinically unacceptable cases, thereby supporting their use as quantitative indicators for DWI image distortion assessment.

\subsection{Performance evaluation of distortion classification models}
Four models were evaluated for predicting DWI geometric distortion severity: segmentation-based model, registration-based model, naive combined model, and ensemble model. The full dataset of 140 cases was randomly divided into a training set (\(n = 85\)) and an independent testing set (\(n = 55\)), with stratification by severity group to maintain class proportions. Model performance was evaluated on the training set using five-fold cross-validation and further assessed on the independent test set. Evaluation metrics included accuracy, recall, specificity, precision, F1-score, and area under the receiver operating characteristic curve (AUC).

\begin{figure}[htbp]
    \centering
    \includegraphics[width=0.9\linewidth]{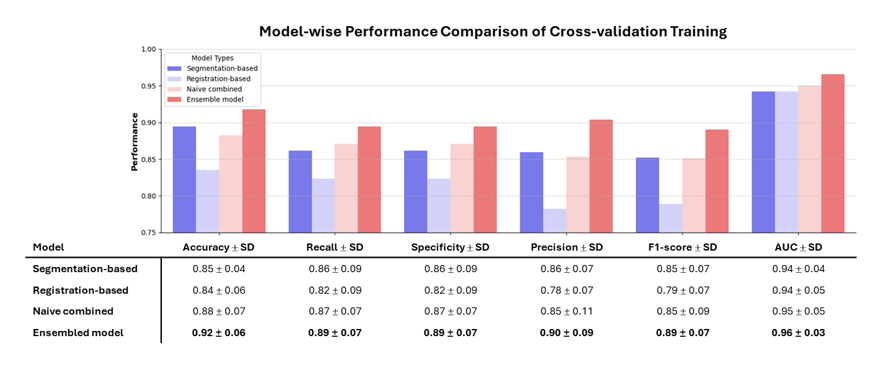}
    \caption{Model-wise performance comparison on five-fold cross-validation training sets. Performance metrics evaluated include accuracy, recall, specificity, precision, F1-score, and area under the receiver operating characteristic curve (AUC). Four model types are compared: the segmentation-based model, the registration-based model, the naive combined model, and the ensemble model. The ensemble model demonstrates superior performance across all evaluated metrics in the training phase. Performance values are presented as mean \(\pm\) standard deviation across folds.}
    \label{fig:sup}
\end{figure}

Figure~\ref{fig:sup} shows the cross-validation performance of all four models on the training set. As detailed in the table, all models were able to differentiate distortion severity groups with accuracy exceeding 80\%. Among the single-method models, the segmentation-based model slightly outperformed the registration-based model. The ensemble model consistently achieved superior performance across all metrics, with the highest mean AUC (\(0.96 \pm 0.03\)) and F1-score (\(0.89 \pm 0.07\)), indicating improved overall classification capability.

\begin{figure}[htbp]
    \centering
    \includegraphics[width=\linewidth]{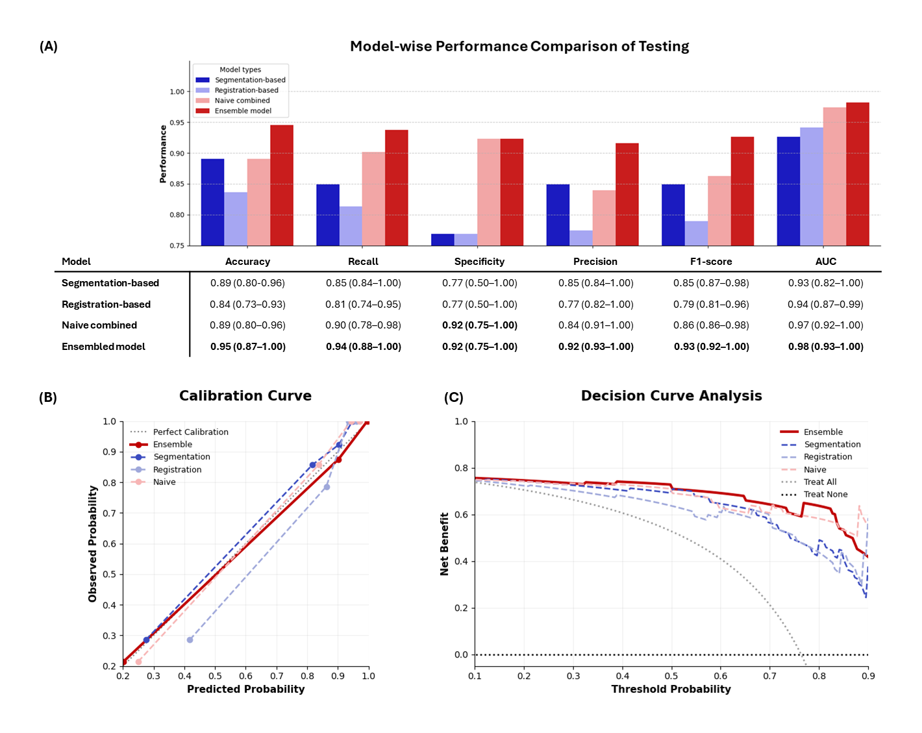}
    \caption{Model-wise performance comparison on independent testing sets. Four model types are compared: the segmentation-based model (blue), the registration-based model (purple), the naive combined model (light red), and the ensemble model (red). (A) Performance metrics evaluated include accuracy, recall, specificity, precision, F1-score, and area under the receiver operating characteristic curve (AUC). Performance values are presented as point estimates with 95\% confidence intervals. The ensemble model demonstrates superior performance across all evaluated metrics. (B) Calibration curves; the diagonal line indicates perfect calibration. (C) Decision curve analysis showing net benefit across threshold probabilities for the models along with treat-all and treat-none strategies.}
    \label{fig:fig6}
\end{figure}

To further assess generalizability, the ensemble model was retrained on the full training set and evaluated on the independent testing set. As shown in Figure~\ref{fig:fig6}(A), the ensemble model demonstrated consistently strong performance across all metrics, with numerically higher accuracy (0.95, 95\% CI: 0.87--1.00), F1-score (0.93, 95\% CI: 0.92--1.00), and AUC (0.98, 95\% CI: 0.93--1.00) compared with the individual models. On the independent test set, the ensemble model resulted in two false-negative and one false-positive classifications, representing fewer misclassifications than the individual segmentation-based, registration-based, and naive combined models.

Calibration analysis showed that the ensemble model demonstrated good agreement between predicted probabilities and observed outcomes, with curves closely approximating the line of perfect calibration (Figure~\ref{fig:fig6}(B)). Decision curve analysis indicated higher net benefit for the ensemble model across a range of clinically relevant threshold probabilities compared with the individual models and treat-all or treat-none strategies (Figure~\ref{fig:fig6}(C)).

These results demonstrate that integrating complementary distortion factors through an ensemble framework enhances model robustness and improves the accuracy of classifying DWI scans as clinically acceptable or unacceptable.

\section{Discussion}
In this study, we proposed AutoIQ for automatic quantification of distortion severity in prostate DWI. Two computational distortion quantification methods were developed and evaluated: one based on anatomical segmentation and the other on image registration. Both approaches demonstrated the feasibility of quantifying geometric distortion in DWI scans. The ensemble model integrates these complementary distortion indicators and translates them into clinically actionable decisions, specifically whether a DWI scan is acceptable or requires repeat acquisition. By providing automated scan-quality assessment, the proposed framework may aid MRI technologists in making timely decisions regarding repeat acquisition and help standardize prostate MRI quality.

The two computational methods can quantify geometric distortion independently and complement each other in assessing distortion in prostate DWI. The segmentation-based method, which relies on accurate delineation of anatomical boundaries, provides accurate distortion factors when prostate segmentation is precise. In most cases, accurate prostate gland segmentation enables the segmentation-based distortion factor to outperform the registration-based method across various metrics. However, for patients with a large volume of rectal gas or metallic implants, segmentation may be inaccurate due to severe distortion and atypical anatomy. In such scenarios, the registration-based method offers a complementary advantage. It uses the segmentation only as a rough approximation for gland localization and is therefore less sensitive to minor segmentation inaccuracies. Nevertheless, it may be affected by over-registration within the gland due to resolution and intensity differences between T2WI and DWI sequences. These characteristics make the two methods naturally complementary. The final ensemble model combines precision from segmentation-based features, global robustness from registration-based features, and joint feature representation from the naive combined model, resulting in improved performance across all evaluation metrics.

Previous studies have proposed methods to quantify geometric distortion in prostate DWI. For example, Gill et al. introduced a contour-based approach that quantified distortion by analyzing boundary displacements between DWI and T2WI in polar coordinates, enabling whole-organ distortion mapping \citep{Gill2017method}. Other studies have quantified distortion using prostate diameter mismatch between DWI and T2WI and investigated how different DWI sequences and acquisition parameters influence distortion levels \citep{Stocker2018Image,Donato2014Geometric}. While these studies provided measurements for image distortion assessment, they were limited to generating distortion maps or numeric scores without integrating their outputs into clinically actionable models. The proposed ensemble framework incorporates two complementary distortion quantification methods and generates a prediction of distortion severity as acceptable or severe. This function may allow for immediate feedback during scan acquisition and support quality-control decisions, addressing a key translation gap in prior work.

PI-QUAL is an established framework for comprehensive prostate MRI quality assessment \citep{Giganti2020PIQUAL}. Formal PI-QUAL scoring was not performed in this study, as the present work focuses specifically on geometric distortion in DWI. By targeting distortion-related failure modes, AutoIQ is positioned to provide immediate feedback during MRI scanning to prompt and guide reacquisition, whereas comprehensive image-quality evaluation frameworks, such as PI-QUAL, are better suited as quality gatekeepers during image interpretation. Nonetheless, because distortion is an important aspect of prostate MRI quality evaluation, examinations categorized as severe distortion by AutoIQ would likely have impaired lesion localization or unreliable ADC interpretation.

This study has several limitations. First, the model was developed retrospectively using data from a single center, which may limit its generalizability across different institutions, scanner platforms, and imaging protocols. Second, although the dataset encompassed a wide range of distortion severity, including severe cases from patients with bilateral implants, the sample size was relatively small. Under this constraint, we developed two computational methods to quantify distortion severity, which effectively distinguished between clinically actionable decision-making categories: acceptable scans requiring no rescan and severe scans requiring repeat acquisition. These distortion factors served as concise and interpretable features for conventional machine learning classifiers.

The current study can be extended in several directions. First, we aim to improve the proposed framework by expanding the dataset to include multicenter, multi-scanner prostate MRI to enhance generalizability. With access to larger and more diverse datasets, we will explore incorporating additional quantitative features, including image-derived texture measures and deep learning-based representations, which may further enhance predictive accuracy. Second, we plan to implement prospective clinical validation by integrating the framework into MRI scanner software. Future integration into MRI acquisition workflows may allow AutoIQ to deliver real-time distortion assessment and support proactive image-quality decisions to help maintain diagnostically interpretable prostate DWI. Furthermore, the framework could be extended to assess distortion in other anatomical regions or used with different MRI sequences, broadening its potential clinical applicability.

\section{Conclusion}
In conclusion, AutoIQ is an ensemble machine learning framework for automatic quantification and classification of geometric distortion severity in prostate DWI. The proposed method integrates two complementary computational algorithms for distortion quantification and achieves strong performance in classifying distortion severity. By providing automated quantification of geometric distortion, the proposed framework may serve as a decision-support tool to assist MRI technologists in identifying scans that require repeat acquisition, promoting the consistent acquisition of clinically acceptable prostate MRI. With future integration into MRI scanner workflows, AutoIQ may enable real-time feedback during acquisition to support image quality control.

\section*{Funding}
This research was supported by Siemens Medical Solutions USA, Inc. and the National Institutes of Health under award R01CA217098.

\section*{Acknowledgments}
The authors thank Jenny Park for patient recruitment and the Cedars-Sinai Research Imaging Core staff for their assistance with MRI scans.

\bibliographystyle{unsrtnat}
\bibliography{references}  

\end{document}